\documentclass[%
reprint,
nofootinbib,
 amsmath,amssymb,
 aps,
 prd
]{revtex4-1} 

\usepackage[utf8]{inputenc}

\usepackage{amsmath}
\usepackage{dcolumn}
\usepackage{bm}
\usepackage{csquotes}
\usepackage{graphicx}
\usepackage[english]{babel}
\usepackage[T1]{fontenc}
\usepackage{amssymb}
\usepackage{setspace}
\usepackage{lmodern}
\usepackage{geometry}
\usepackage{enumerate}
\usepackage{caption}
\usepackage{subcaption}
\usepackage{algorithm}
\usepackage[]{algpseudocode}
\usepackage[dvipsnames]{xcolor}
\usepackage[normalem]{ulem}
\usepackage{hyperref}
\hypersetup{
    colorlinks,
    citecolor=blue,
    filecolor=blue,
    linkcolor=blue,
    urlcolor=blue
} 
\usepackage{soul}

\def\lg{\lambda_g}

\usepackage{natbib}
\begin{document}

\preprint{APS/123-QED}

\title{Bayesian test of the mass of the graviton with planetary ephemerides}

\author{Vincenzo \surname{Mariani}}
 \email{vmariani@geoazur.unice.fr}
\affiliation{%
G\'eoazur, Universit\'e C\^ote d'Azur, Observatoire de la C\^ote d'Azur, CNRS, 250 av. A. Einstein, 06250 Valbonne, France}
\author{Agn\`es  \surname{Fienga}}%
 \email{agnes.fienga@oca.eu}
\affiliation{%
G\'eoazur, Universit\'e C\^ote d'Azur, Observatoire de la C\^ote d'Azur, CNRS, 250 av. A. Einstein, 06250 Valbonne, France,
IMCCE, Observatoire de Paris, PSL University, CNRS, 77 av. Denfert-Rochereau, 75014 Paris, France}%

\author{Olivier \surname{Minazzoli}}
 \email{ominazzoli@gmail.com}
\affiliation{
Artemis, Universit\'e C\^ote d'Azur, Observatoire de la C\^ote d'Azur, CNRS, BP4229, 06304, Nice Cedex 4, France,
Bureau des Affaires Spatiales, 2 rue du Gabian, 98000 Monaco}
\author{Micka\"{e}l \surname{Gastineau}}%
 \email{mickael.gastineau@obspm.fr}
\affiliation{%
IMCCE, Observatoire de Paris, PSL University, CNRS, 77 av. Denfert-Rochereau, 75014 Paris, France}%

\author{Jacques \surname{Laskar}}%
 \email{jacques.laskar@obspm.fr}
\affiliation{%
IMCCE, Observatoire de Paris, PSL University, CNRS, 77 av. Denfert-Rochereau, 75014 Paris, France}%

\begin{abstract}

In this work, we investigated Bayesian methodologies for constraining in the Solar System a Yukawa suppression of the Newtonian potential---which we interpret as the effect of a non-null graviton mass---by considering its impact on planetary orbits. Complementary to the previous results obtained with INPOP planetary ephemerides, we consider here a Markov Chain Monte Carlo approach associated with a Gaussian Process Regression for improving the resolution of the constraints driven by planetary ephemerides on the graviton mass in the Solar System.
At the end of the procedure, a posterior for the mass of the graviton is presented, providing an upper bound at $1.01 \times 10^{-24} \; eV c^{-2}$ (resp. $\lambda_g \geq 122.48 \times 10^{13} \; km$) with a $99.7\%$ confidence level. The threshold value represents an improvement of 1 order of magnitude relative to the previous estimations. This updated determination of the upper bound is mainly due to the Bayesian methodology, although the use of new planetary ephemerides  (INPOP21a used here versus INPOP19a used previously) already induces a gain of a factor 3 with respect to the previous limit. The INPOP21a ephemerides is characterized by the addition of new Juno and Mars orbiter data, but also by a better Solar System modeling, with notably a more realistic model of the Kuiper belt.
Finally, by testing the sensitivity of our results to the choice of the \emph{a priori} distribution of the graviton mass, it turns out that the selection of a prior more favorable to zero-mass graviton (that is, here, General Relativity) seems to be more supported by the observations than non-zero mass graviton, leading to a possible conclusion that planetary ephemerides are more likely to favor  General Relativity.
\end{abstract}

\maketitle

\date{\today}
\section{Introduction}
\label{sec:introduction}
Planetary ephemerides have evolved with the {observational} accuracy obtained for the astrometry of planets and natural satellites thanks to the navigation tracking of spacecraft (s/c) orbiting these systems. Since the late XIXth century the astrometry of planets has known a significant improvement leading to an increased accuracy of the dynamical theories describing their motions. 
The motion of the planets and asteroids in our Solar System can be solved directly by the numerical integration of their equations of motion.
The improved (present and future) accuracy in the measurements of observables from space missions like Cassini–Huygens, Mars Express (MEX), Venus Express (VEX), BepiColombo etc. makes the Solar System a suitable arena to test General Relativity Theory (GRT) as well as alternative theories of gravity by mean of Solar System ephemerides.  
INPOP planetary ephemerides are developed since 2003, integrating numerically the Einstein-Infeld-Hoffmann equations of motion proposed by \cite{moyer1971mathematical, moyer2003monography},  and fitting the parameters of the dynamical model to the most accurate planetary observations following \cite{Soffel_2003}.
Testing alternative theories of gravity with planetary ephemerides  consist in changing the metric of GRT into alternative frameworks and consequently modifying the equations of
motion, the light time computation and the definition of time-scales used for the construction of planetary ephemerides {(see \cite{LRR22} for a review)}.  In principle, such modifications of GRT can be {summarized as} considering additional terms to GRT fundamental equations,
such as {a Yukawa suppression of the Newtonian potential}.
At the phenomenological level, a mass of the gravitational interaction\footnote{{Reported as \textit{the mass of the graviton} in the rest of the paper, for convenience.}} is often assumed to either lead to a modification of the dispersion relation of gravitational-waves \cite{LIGOVirgo_testsGRT_12_2021} or to lead to a Yukawa suppression of the Newtonian potential\footnote{{For more information on the status of current theoretical models of massive gravity, we refer the reader to \cite{RevModPhys.89.025004, derham:2014lr}}.} \cite{will:1998pd}. 
%
%
Recently\footnote{{Following his more-than-twenty years old seminal work \cite{will:1998pd}.}} in \cite{will:2018cq},
Will argued that Solar System observations and planetary dynamics could be used to improve the constraints on the mass of the graviton $m_g$.
However Will used results based on statistics of postfit residuals of the Solar System ephemerides that are performed without including the effect of a massive graviton inside the equations of motion. In order to overcome the consistency issues that are raised by such type of analyses---that is, which are based on postfit residuals---
we investigate an original approach that is based on a statistical inference of the mass of the graviton $m_g$ within the framework of INPOP as presented in \citep{bernus2019, Bernus2020}.

In this work, the approach we decided to use is partially Bayesian and several tools like Markov Chain Monte Carlo (MCMC), Gaussian Process and Bayes Factor are employed, in order to exploit at the best the values of INPOP $\chi^2$. 
Contrarily to \cite{Bernus2020}, {a full posterior distribution of the graviton mass $m_g$ is deduced, using a Metropolis-Hastings algorithm which optimizes the information contained in such a distribution, and a Gaussian Process Regression which improves the resolution of the search process.} 
%
%
{Moreover}, by considering two different assumptions for the a priori distribution of the graviton mass $m_g$, we are able i) with an uniform distribution, to give a new upper limit for $m_g$, improving the limit by a factor 30 relative to \cite{Bernus2020} (see sections \ref{sec:results_flat_wo_uncertainty} and \ref{sec:results_flat_and_uncertainty}) ii) with a Laplace distribution, to demonstrate that, with the present accuracy of the planetary ephemerides, GRT is sufficient for explaining the observations (see section \ref{sec:resultslaplace}). 
{In other words} we can infer that the nowadays planetary ephemerides tend to prefer a non-massive graviton. In Appendix \ref{sec:method}, we give more details on the methods applied.
\section{Theoretical framework}

\subsection{Massive graviton phenomenology}
\label{sec:graviton}
There is not an unique definition of what a massive gravity may mean \citep{derham:2014lr}. Massive interactions usually lead to a Yukawa suppression of those interactions on the scale of  the Compton wavelength, $\lg$. But gravity is different, and therefore it may not be the case for a fully consistent theory of massive gravity \citep{derham:2014lr}. Nevertheless, from a phenomenological point of view \cite{will:1998pd}, one can test whether or not there is a Yukawa suppression of the gravitational potentials at the level of the Solar System ---see, e.g., \cite{will:2018cq,RevModPhys.89.025004} and references therein. Let us note that, while one often talks about a "graviton mass" in the literature \cite{will:2018cq,RevModPhys.89.025004}, the word "graviton" is mostly used for convenience since everything is considered at the classical level only. (The same way that some may use the word "photon" for classical electrodynamics). Formally, this would lead to the following modification of the Newtonian potential $w_{\textrm{Newton}}$ \citep{will:2018cq,RevModPhys.89.025004}
\begin{equation}
w=w_{\textrm{Newton}}\mathrm{exp}(-r/\lg),
\label{eq:massive}
\end{equation}
which can be developed as \cite{bernus2019}
\begin{equation}\label{eq:massivepert}
w=w_{\textrm{Newton}} \left(1+\frac{1}{2} \frac{r^2}{\lg^2} \right)+\mathcal{O}(\lg^{-3}),
\end{equation}
after a convenient change of coordinate system that absorbs the constant term in the gravitational potential, and which has no impact on the observables. By analogy with standard quantum physics, the Compton length can also be interpreted in terms of {a} mass of the graviton $m_g$ following the relation:
\begin{equation}\label{eq:mglg}
\lg=\frac{\hbar}{c m_g},
\end{equation}
with $\hbar$ the Planck constant, and $c$ the speed of light. (In some sense this is also one of the reasons why, beyond simple convenience, one often talks about a "graviton mass" despite working at the classical level, since the translation of the Yukawa suppression length in terms of a mass involves the quantum of action $\hbar$). In that situation, the equation of motion has only one extra term with respect to the usual EIHDL equations 
that reads 
\begin{equation}
 \delta a^i = \frac{1}{2} \sum_{P} \mu_P \frac{  c^2 m_g^2 }{\hbar^2} \frac{x^i - x^i_P}{r} + \mathcal{O}(m_g^{3}), \label{eq:accgravmas}
\end{equation}
where $\mu_P$ is the gravitational parameter $\mu_P = G M_P$. Further assuming that light still propagates along null geodesics, the Shapiro delay reads 
\begin{equation}\label{eq:shapiro_lg}
\begin{split}
c (t_r-t_e) & =c(t_r-t_e)_{GRT} \\
+ & \sum_{A}\frac{\mu_A}{c^2}\frac{c^2 m_g^2}{2 \hbar^2} \ln \biggl[  b^{2}\frac{\bm{n}\cdot\bm{r}_{rA}+r_{rA}}{\bm{n}\cdot\bm{r}_{eA}+r_{eA}}  \\
& + \bm{n}\cdot (r_{rA}\bm{r}_{rA}-r_{eA}\bm{r}_{eA})  \biggr]  \\ &+ \mathcal{O}(m_g^{3}). 
\end{split}
\end{equation}
The term $c(t_r-t_e)_{GRT}$ corresponds to the GRT light time, 
$b$ is the minimal distance between the light path and the central body (here the Sun). This expression is an approximation at the {c$^{-2}$} level. Let us note, however, that the correction to the Shapiro delay due to the graviton mass has been found to be negligible in practice as it was discussed in \cite{bernus2019} using the Compton wavelength formalism.

In  \cite{bernus2019} and \cite{Bernus2020}, planetary ephemerides have been fully developed in the massive gravity framework of Eqs. (\ref{eq:accgravmas}-\ref{eq:shapiro_lg}) using the equivalent Compton wavelength formalism and fitted over the data sample for INPOP17a and INPOP19a respectively. The results of these investigations are gathered in \cite{LRR22}.

It is somewhat interesting to compare these constraints to the ones deduced from the observation of gravitational waves \cite{will:1998pd,will:2018cq}. Indeed, one can assume that a massive gravitational field that leads to Eq. (\ref{eq:massive}) might also modify the dispersion relation of gravitational waves as follows \citep{GWTC3b,will:1998pd}
\begin{equation}
E^{2}=p^{2} c^{2}+m_g c^2, \label{eq:dispm}
\end{equation}
where $E$ and $p$ are the energy and momentum of the wave.
Such a modified dispersion relation causes gravitational wave frequency modes to propagate at different speeds, leading to an overall modification of the phase morphology of gravitational waves with respect to the GRT predictions. Since the morphology of gravitational wave phase has been consistent with General Relativity so far, it led to severe constraints on the value of the graviton mass: $m_g \leq 1.27 \times 10^{-23} eVc^{-2}$ at 90\% confidence level \cite{LIGOVirgo_testsGRT_12_2021}, whereas previous results with ephemerides were at the few $10^{-23} eVc^{-2}$ level \cite{Bernus2020}. 

Each type of constraints is relevant in its own right given that they test different phenomenologies, which may (or may not) be related, depending on the underlying massive gravity theory that one is considering. For instance, screening mechanisms may suppress one effect but not the other \citep{derham:2014lr}.

\subsection{Planetary ephemerides construction}
\label{sec:methodinpop}

INPOP (Int\'egrateur Num\'erique Plan\'etaire de l’Observatoire de Paris) is a planetary ephemerides that is built by integrating numerically the equations of motion of the Solar System objects following the formulation of Moyer \cite{moyer2003monography}, and by adjusting to Solar System observations such as {space mission navigation and radio science data, ground-based optical observations or lunar laser ranging} (\cite{Fienga2008, INPOP21a}).
In addition to adjusting the astronomical intrinsic parameters, it can be used to constrain parameters that encode deviations from GRT \cite{inpop10a_ff, verma2014, Fienga-2015, Viswanathan2018}, such as the Compton wavelength $\lambda_g$ as defined in Eq. \eqref{eq:mglg}.
As long as $m_g$ is small enough, the gravitational phenomenology in the Newtonian regime {recovers the} one of GRT. \\

%
But, as discussed in \cite{will:2018cq},  a graviton mass would indeed lead to a modification of the perihelion advance of Solar System bodies. Based on published constraints on the perihelion advance of Mars---or on the post-Newtonian parameters $\gamma$ and $\beta$---derived from Mars Reconnaissance Orbiter (MRO) data, Will had estimated that the graviton  mass should be smaller or equal to  $(4 - 8) \times 10^{-24} \; eVc^{-2}$ depending on the specific analysis.
But, as an input for his analysis, Will uses results based on interpreting statistics of postfit residuals of the Solar System ephemerides obtained in various frameworks (PPN and GRT) as possible outcome of the graviton influence.
However, first of all---unlike the historical occurrence of the substantial error in the perihelion advance of Mercury computed in Newton's theory---a lot of different contributions from the details of the Solar System model being used could explain the rather small postfit differences between computed and observed positions \cite{LRR22}.
Furthermore various parameters of the ephemerides (e.g., masses, semimajor axes, etc.) are  more or less correlated to $m_g$ as it is shown in \cite{bernus2019}. 
Therefore any kind of signal introduced by $m_g > 0$ can, in part, be reabsorbed during the fit of other correlated parameters. In order to {overcome the correlation issues decribed previously,} we investigate a new approach based on a statistical inference on the mass of the graviton $m_g$ within the full framework of INPOP. 
Planetary ephemerides are developed in the framework described in Sec. \ref{sec:graviton} where planetary equations of motion but also Shapiro delay are modified according to Eqs. (\ref{eq:accgravmas}) and (\ref{eq:shapiro_lg}).

In this work, we will use the INPOP21a planetary ephemerides \cite{INPOP21a} that  benefits from the latest Juno and Mars orbiter tracking data up to 2020 as well as a fit of the Moon-Earth system to LLR observations also up to 2020. For a more detailed review about this specific version, the reader can refer to \cite{INPOP21a} whereas \cite{2022IAUS..364...31F} gives more descriptions regarding recent GRT tests obtained with INPOP21a and INPOP19a. INPOP21a is more accurate than INPOP19a especially for Jupiter and Saturn orbits as additional Juno observations of Jupiter were used covering a 4-years period when only 2.5 years were considered in INPOP19a. Consequently, a more realistic model of the Kuiper belt was implemented in INPOP21a, leading to an improvement of about 1.4 on the Jupiter orbit accuracy. Additionally, 2 years of Mars Express navigation data have been added to the 13 years already implemented in INPOP19a. This increase of Mars orbiter data improves mainly the stability of the ephemerides and its extrapolation capabilities \cite{10.1093/mnras/stz3407}. In terms of adjustment, in addition to the initial conditions of the planetary orbit,  the gravitational mass of the Sun, its oblateness and the ratio between the mass of the Earth and the one of the Moon, 343 asteroid masses are fitted in INPOP21a following the procedure described, for example, in \cite{10.1093/mnras/stz3407}. A mass representing the average effect of 500 trans-neptunian objects has also been added as described in \cite{INPOP21a}. A total of 401 parameters are accounted for the INPOP21a construction. They constitute the list of astronomical parameters we will refer to in the following.

\section{Methodology and results}
\label{sec:methods_and_results}

Starting from \cite{Fienga-2015}, as well in \cite{DiRuscioFienga2020, Fienga2020P9, Fienga2016P9}, several tools to assess the goodness of the INPOP fit with respect to modifications {in the equations of motion or in the global framework of the ephemeris} or observations have been used. In particular the computation of the INPOP $\chi^2$ plays an essential role to determine which data or which model improve {significantly} the ephemeris computation. In our work,  as in \cite{Bernus2020}, the computation of $\chi^2(m_g)$ is the output, for a given value of $m_g$, of the INPOP {iterative} fit, after adjustment of all its astronomical parameters (see Sec. \ref{sec:methodinpop}). 
The $\chi^2(m_g)$ is computed following 
\begin{equation}\label{normalized_chi2_form1}
\scalebox{0.95}{$ \chi^2(m_g, \mathbf{k}) \equiv \frac{1}{N_{\text{obs}}} \sum_{i=1}^{N_{\text{obs}}} \left( \frac{g^i(m_g, \mathbf{k}) - d^i_{\text{obs}}}{\sigma_i} \right)^2$ }
\end{equation}
where {$m_g$ is a fixed value, $\mathbf{k}$ are the astronomical parameters fitted with INPOP (see Step 1 on Fig. \ref{fig:block_diagram_pipeline}, Sec. \ref{sec:nutshell} and Sec. \ref{sec:methodinpop}), $N_{\text{obs}}$ is the number of observations, the function $g^i$ represents the computation of observables, the vector $\mathbf{d}_{\text{obs}}=(d^i_{\text{obs}})_i$ is the vector of observations and $\sigma_i$ are the observational uncertainties.}

\subsection{Methodology in a nutshell}
\label{sec:nutshell}

\begin{figure*}
\includegraphics[width=\textwidth]{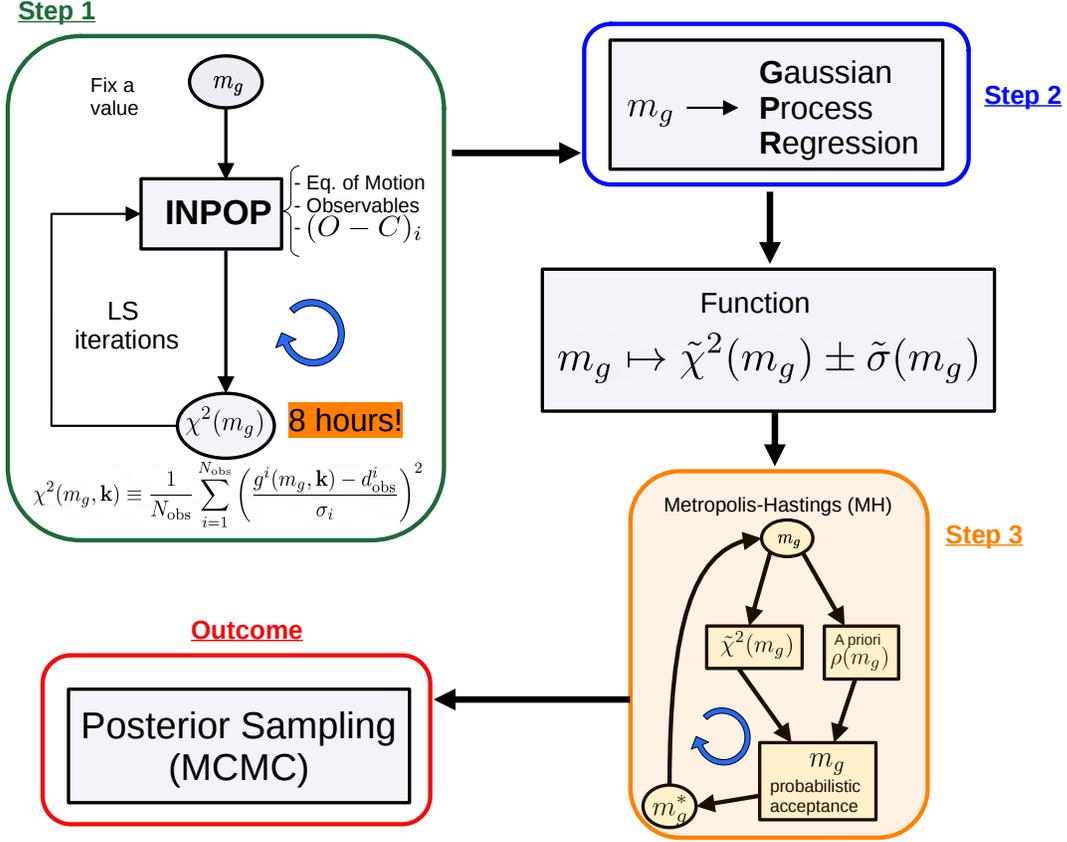}
\caption{General pipeline used to obtain the posterior of $m_g$. The first step is the computation of $\chi^2(m_g)$ for $m_g$ spread over the domain of our interest. Whereupon a GPR is computed, obtaining the interpolation $m_g \longmapsto \tilde{\chi}^2(m_g)$ with the corresponding uncertainty.  The outcome is the posterior for the mass of the graviton $m_g$. The fully detailed description of method and pipeline can be found in \cite{2023arXiv230305298M}.}
\label{fig:block_diagram_pipeline}
\end{figure*}

As said in Sec. \ref{sec:introduction}, we want to obtain a posterior for the mass of the graviton $m_g$. The general pipeline used to obtain such posteriors is summarized on Fig. \ref{fig:block_diagram_pipeline} and works as follows.
First (Step 1 on Fig. \ref{fig:block_diagram_pipeline}), we compute the value of $\chi^2(m_g)$ (see Eq. \eqref{normalized_chi2_form1}) for several different values of $m_g$ spreading over the domain of our interest. For a given $m_g$ the value $\chi^2(m_g)$ is obtained as the outcome of the full INPOP iterative adjustment letting fixed $m_g$. In such a fit, the astronomical parameters $\mathbf{k}$ of Eq. \eqref{normalized_chi2_form1} are adjusted  with the least squares procedure, whereas $m_g$ is a fixed value. This $\chi^2$ estimation is necessary since it ensures the $m_g$ contribution to the dynamics considered, avoiding high correlations between $m_g$ and the other astronomical parameters $\mathbf{k}$ (see \cite{bernus2019}).
The iterative fit can take up to 8 hours of computation for one given $m_g$ value. But, in order to use MCMC algorithm, it is necessary to evaluate {\it sequentially} the likelihood, and so the $\chi^2$, for thousands of different values of $m_g$ (see Appendix \ref{sec:method}). In this context, the direct $\chi^2$ computation becomes difficult. 
{In order to overcome the problem of the computation time, the second step of the method (Step 2 on Fig. \ref{fig:block_diagram_pipeline}) is the use of } a Gaussian Process Regression (GPR) to interpolate among the values $(m_g, \chi^2(m_g))$, already computed during the Step 1. Starting from this set of points we obtain with the GPR a function $m_g \mapsto \tilde{\chi}^2(m_g)$ which is the interpolation of that set of points, along with an uncertainty $\tilde{\sigma}(m_g)$ relative to the possible error of interpolation in $\tilde{\chi}^2(m_g)$. The interpolation $m_g \mapsto \tilde{\chi}^2(m_g)$ is necessary to run the Metropolis-Hasting  (MH) algorithm {(Step 3 on Fig. \ref{fig:block_diagram_pipeline})}, which has, as an outcome, the posterior density for $m_g$ in the form of a Markov Chain Monte Carlo (MCMC).
The Step 3 of the method is then the  Metropolis-Hasting  (MH) algorithm for selecting possible values of $m_g$. The algorithm is described in Appendix \ref{sec:methodMH} and more details can be found in \cite{2023arXiv230305298M}. In the MH algorithm, a prior density distribution has to be settled for the parameter to sample with the MCMC. 
For this work, {we used two different prior density distributions} of the graviton mass as inputs for the MH algorithm in order to test {the sensitivity of the posterior to the choice of the prior and} if there is any gain in {considering} a non-zero mass graviton (with an uniform prior distribution, see Sections \ref{sec:results_flat_wo_uncertainty} and \ref{sec:results_flat_and_uncertainty}) versus a massless graviton (with a half-Laplace prior distribution, see section \ref{sec:resultslaplace}).

\subsection{{Results with the uniform prior distribution and upper bound}}
\label{sec:results_flat_wo_uncertainty}

In terms of prior for the mass of the graviton, we firstly chose an uniform distribution with large intervals of values encompassing the latest results from \cite{Bernus2020} but without giving preference to any possible values.
Following the step 2 of Fig. \ref{fig:block_diagram_pipeline}, for each value of the graviton mass proposed by the MH algorithm, {the value $\tilde{\chi}^2(m_g)$ is computed instead of $\chi^2(m_g)$.} 
Based on a first grid of fully converged runs, we use { the Gaussian Process Regression as a form of interpolation} to increase the resolution of the MH algorithm, along with GPR error estimations to assess the uncertainties of such regression. Details about GPR and GPR error estimations are given in \cite{2023arXiv230305298M}.
\subsubsection{MH algorithm and Gaussian Process Regression (GPR)}
\label{sec:resultsflatMH}
\begin{figure}[ht]
\centering
  \includegraphics[width=\linewidth]{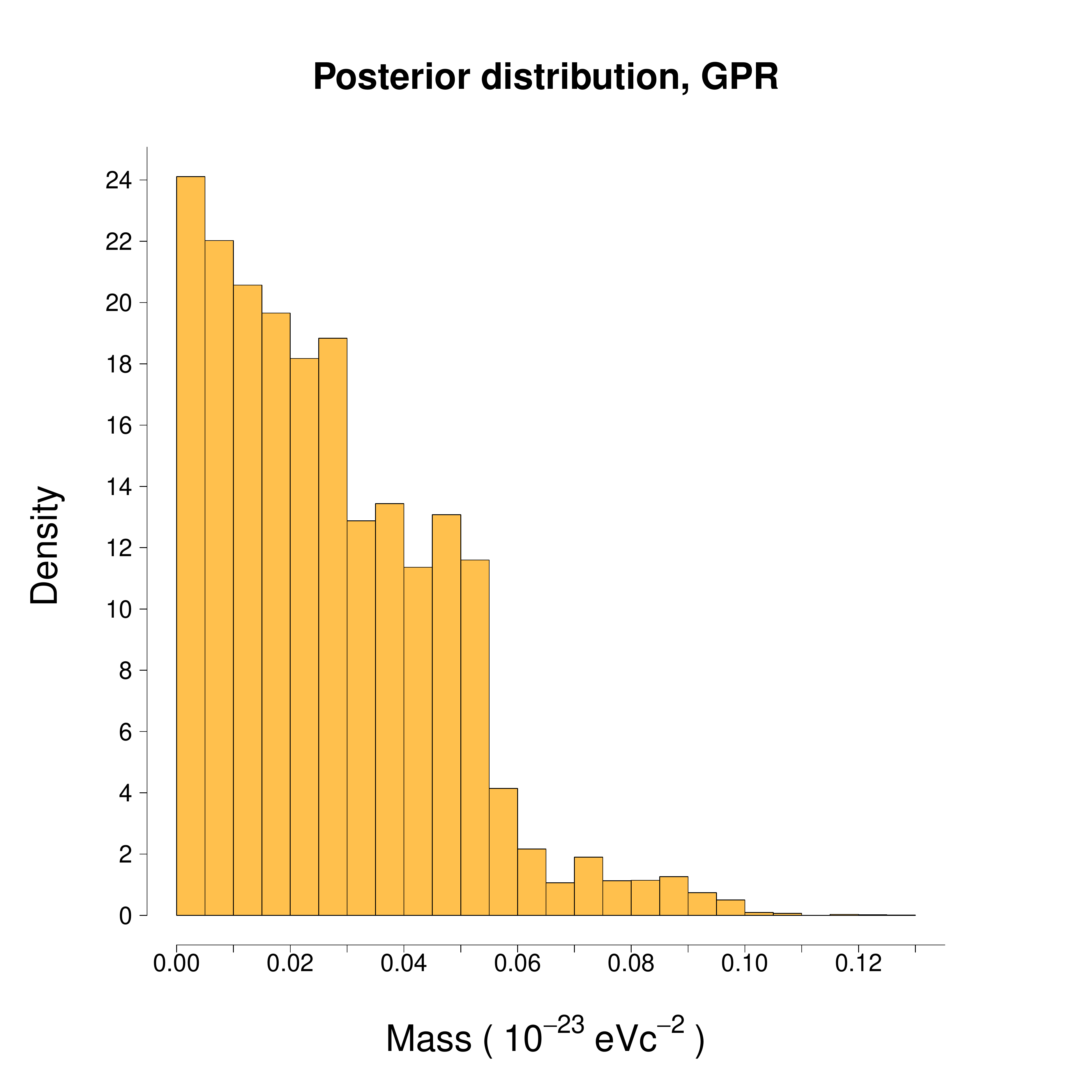}
   \caption{Density for the posterior probability distribution used as target probability. {The prior was a {uniform} prior between $0$ and $3.62 \times 10^{-23} \; eVc^{-2}$.}}
    \label{fig:55_iter_GPR_histogram_joint}
\end{figure}

On Fig. \ref{fig:55_iter_GPR_histogram_joint} is plotted the posterior density of probability obtained with MH algorithm associated with GPR, supposing an uniform prior. 
Contrarily to a detection curve --- which in our case would look close to a Gaussian-like posterior centered on a positive value (see i.e. \cite{2023arXiv230305298M}) --- from Fig. \ref{fig:55_iter_GPR_histogram_joint}, we do not have a single figure that we could choose as value for the mass of the graviton. Indeed the shape is not a \emph{bell}{, nor does it show} an individual peak. 
The quantile at $97 \%$ is $0.0985 \times 10^{-23} \; eVc^{-2}$. 

The posterior plotted on Fig. \ref{fig:55_iter_GPR_histogram_joint} also tends to concentrate close to $m_g=0$ with decreasing steps for larger $m_g$ up to $m_g < 0.15 \times 10^{-23} \; eV c^{-2}$. 
The conclusion drawn is that, within the GPR approximation of $\chi^2$ we do not have any detection for $m_g \neq 0$. 
Thus we can't provide an estimated value for the mass of the graviton, but we can give a 99.7$\%$ upper limit as quantile of the deduced mass posterior, that is $m_g \leq 0.98 \times 10^{-24} \; eV c^{-2}$.

\subsubsection{Uncertainty Assessment}
\label{sec:results_flat_and_uncertainty}

\begin{figure*}
    \includegraphics[width=.55\linewidth]{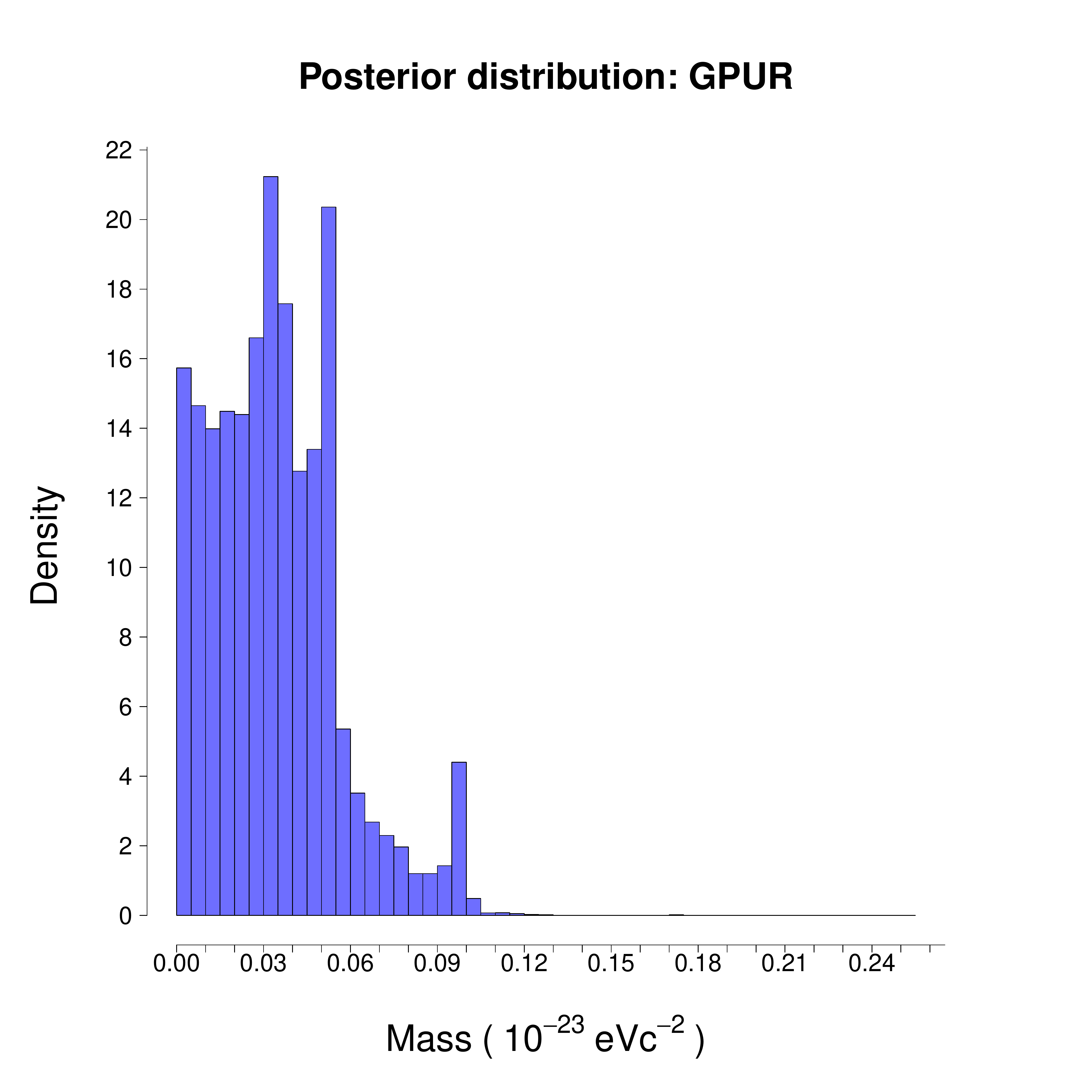}\includegraphics[width=.55\linewidth]{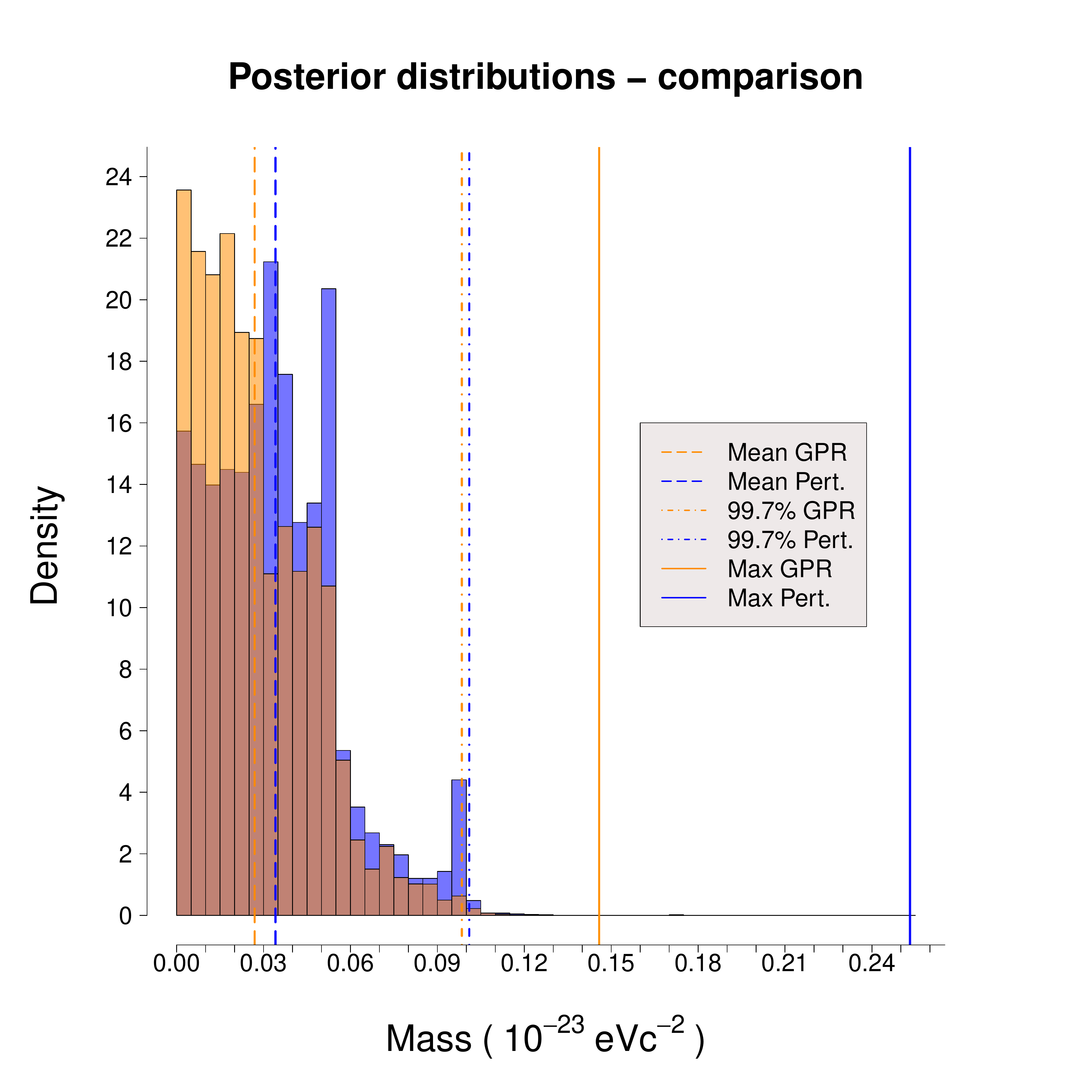}
    \caption{{Posterior probability distributions obtained from GPR (in orange) and the posterior with GPR error assessment (GPUR left-hand slide and in blue). The dashed lines  represent the averages of the posterior with GPR and GPUR (respectively orange and blue). The dot-and-dashed lines represent the $99.7 \%$ quantiles of the two densities. The solid lines are instead in place of the maximum $m_g$ for each one of the two densities. The brown area of the histogram represents the overlaid zone between the two posteriors presented.}}
    \label{fig:55_iter_300Pert_GPR_histogram_comparison}
\end{figure*}

In order to explore the uncertainties induced by the GPR interpolation on $\tilde{\chi}^2(m_g)$ values, we ran 300 MCMC using MH algorithm. Each MH is performed on a different Gaussian Process Uncertainty Estimation, GPUE (see the black line on Fig. \ref{fig:GPR_only_plot_IT55_zoom3_nogrid_pert}). 
{One GPUE is a perturbation of the GPR within the uncertainty provided by the GPR itself.} The GPUE interpolates { possible ${\chi}^2$ deviations from $\tilde{\chi}^2$ (see \cite{2023arXiv230305298M} for a detailed description on how to encompass uncertainty)} exploring the space of uncertainties in the GP regression. {Running the MCMC on the GPUE is a way to propagate the uncertainty induced by the GPR in the posterior sampling}. We, therefore, built a posterior distribution presented in blue on Fig. \ref{fig:55_iter_300Pert_GPR_histogram_comparison} accounting for the GPR uncertainty, {and we call such a posterior Gaussian Process Uncertainty Realization (GPUR).}

The GPUR result presented on Fig. \ref{fig:55_iter_300Pert_GPR_histogram_comparison} is consistent with the nominal case (see Section \ref{sec:resultsflatMH}), and it shows the posterior with a slightly larger interval of masses. This is actually what it is expected, since the uncertainty of GPR in the present case is small, but not absent. On Fig. \ref{fig:55_iter_300Pert_GPR_histogram_comparison} it is easy to see how much the maximum value of $m_g$ is shifted towards larger $m_g$, passing from the MCMC with GPR to the MCMC on GPUEs.
In particular the average going from $0.26 \times 10^{-24} \; eVc^{-2}$ to $0.34 \times 10^{-24} \; eVc^{-2}$ whereas the maximum mass in the posterior going from $1.45 \times 10^{-24} \; eVc^{-2}$ to $2.53 \times 10^{-24} \; eVc^{-2}$. 
The strategy we propose in {Appendix \ref{sec:method}} relies on the assumption that if we compute the real values of $\chi^2(m_g)$, then we can {estimate $\chi^2$ values in} the zones of domain for which $\chi^2(m_g)$ is unknown, with an uncertainty based on $\chi^2$ values already computed.
In our specific case, the strategy looks like consistent. The outcome of the 300 MCMC runs on different GPUEs is similar with respect to the nominal GPR case with slight differences (see e.g. Table \ref{tab:GPR_300iterations_comparison_55it} and Fig. \ref{fig:55_iter_300Pert_GPR_histogram_comparison}). 
As previously indicated {(see Sec. \ref{sec:resultsflatMH})}, again the GPUR posterior is not similar to the one obtained with a positive detection. 
We are however able to provide limits for the mass.
The upper bound for GPUR we would provide at $99.7\%$ C.L. is $m_g \leq 1.01 \times 10^{-24} \; eV c^{-2}$. This represents an improvement of about 1 order of magnitude from the previous estimations in terms {of upper limit provided for $m_g$ at $99.7 \%$ C.L. . Such a limit is taken from the GPUR posterior (see Fig. \ref{fig:55_iter_300Pert_GPR_histogram_comparison}) since it takes into consideration also the GPR uncertainty.}

\begin{center}
\begin{table}[!ht]
\centering
\caption{\label{tab:GPR_300iterations_comparison_55it} Summary of the outcome for MCMC based on GPR and the posterior GPUR. These values are plotted as vertical lines on Fig. \ref{fig:55_iter_300Pert_GPR_histogram_comparison}. The unit is $10^{-24} eVc^{-2}$. The prior was a {uniform} prior between $0$ and $3.62 \times 10^{-23} \; eVc^{-2}$.}
\begin{tabular}{| c || c | c | c | c | c | c | }
\hline
  &  $ \langle m_g \rangle $ & $99.7 \% $ quantile & $\max\{ m_g \} $  \\ [1.2ex]
\hline \hline 
 
GPR & $ 0.26 $ & $ 0.98 $ &  $1.45$  \\ \hline
GPUR & $ 0.34 $ & $ 1.01 $ &  $2.53 $  \\ \hline

\end{tabular}
\end{table}
\end{center}

Finally, as stressed in Table \ref{tab:GPR_300iterations_comparison_55it}, from a very large uniform prior between $0$ and $3.62 \times 10^{-23} \; eVc^{-2}$, the MCMC algorithm indicates a posterior between $0$ and a maximum of {$2.53 \times 10^{-24} \; eVc^{-2}$}, inducing a significant improvement also  on the possible maximum value for the mass of the graviton.

\subsection{Results with the {half-Laplace} prior distribution}
\label{sec:resultslaplace}

\begin{figure*}
    \centering
    \includegraphics[width=.89\linewidth]{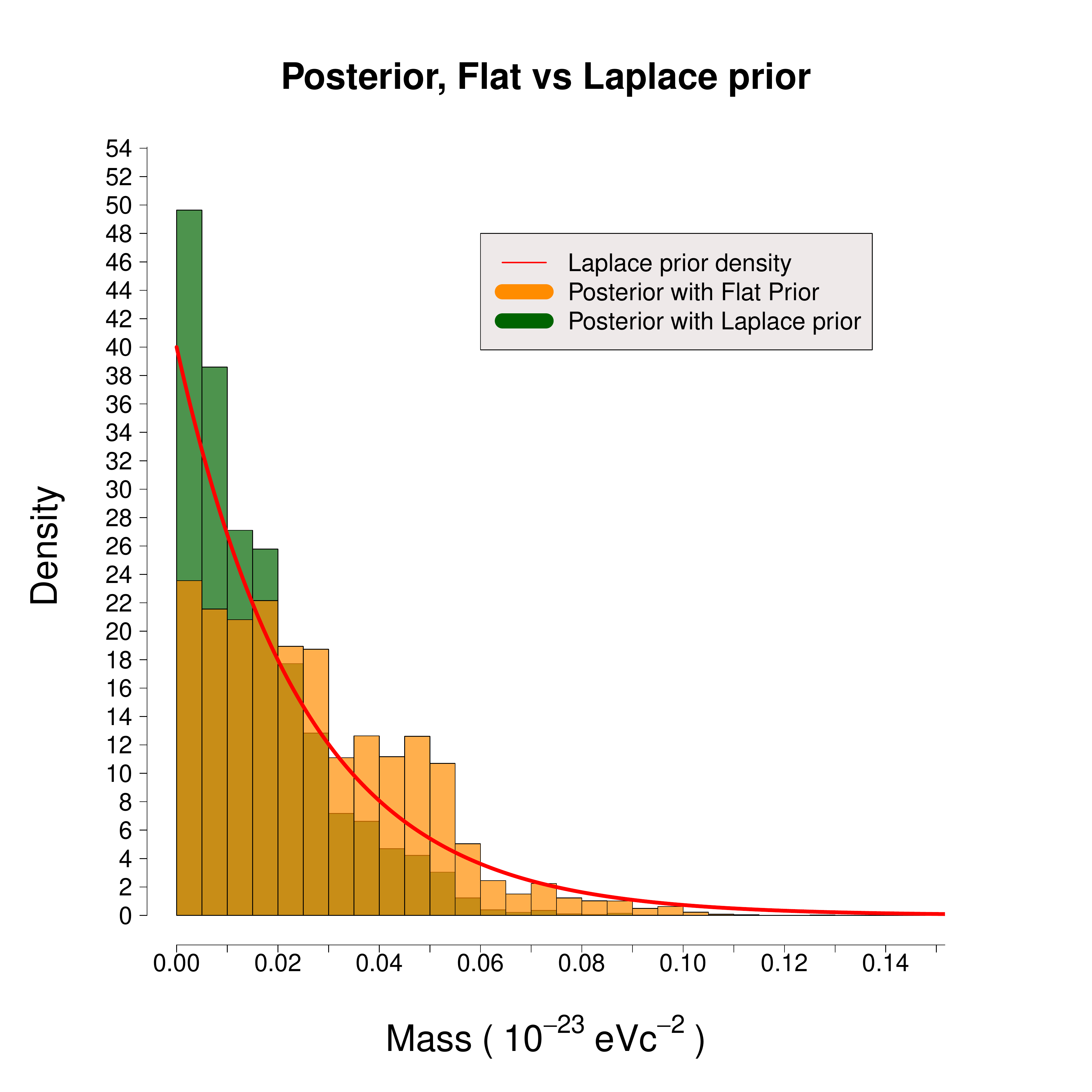}
    \caption{Densities for the posterior probability distributions obtained from GPR with an uniform (flat) prior (in orange) and with a half-Laplace prior (dark green). In red, the shape of the half-Laplace prior used. The brown area is the overlaid zone of the two posteriors}
    \label{fig:55_iter_Flat_Laplace_priors_comparison}
\end{figure*}

The {absence of} {positive} detection can be interpreted in two ways: either because the data employed are not sensitive enough, or because GRT is sufficient to explain the data. {The lack of detection could also depend on the hypothesis made on our a priori knowledge of the mass of the graviton}. In order to discriminate between the former and the latter {and to test the sensitivity of our results to the prior}, we made {an additional experiment}, changing the prior density for $m_g$ in the MH algorithm {from an uniform prior to a half-Laplace distribution}, which gives a prior preference to small graviton masses. We run the MH algorithm again, with the same GPR, by using a half-Laplace prior (red line on Fig. \ref{fig:55_iter_Flat_Laplace_priors_comparison}).
The half-Laplace prior is chosen such that the zone of higher probability of the half-Laplace density shares the same domain of the posterior obtained with the uniform prior. 
The underlying idea of half-Laplace prior is to give preference to the mass value $m_g=0$, being in our case representative of GRT. Doing so we can discern whether the data have enough information to flatten this new prior or not. As one can see on Fig. \ref{fig:55_iter_Flat_Laplace_priors_comparison}, the MH algorithm does not provide the same outcome with the two different priors, and in particular the new posterior (green on Fig. \ref{fig:55_iter_Flat_Laplace_priors_comparison}) turns out to be piled up around $m_g=0$. 
For a validation of the MCMC convergence in the case of half-Laplace prior we refer to \cite{2023arXiv230305298M}.

\section{Discussion}
\label{sec:discussion}

\subsection{Towards a non detection}
\label{sec:discussionNo}

\subsubsection{Non detection with half-Laplace prior}
\label{sec:disc_laplace_prior}
Comparing the outcome obtained with the two different priors (uniform and half-Laplace), we may see whether the new posterior (green on Fig. \ref{fig:55_iter_Flat_Laplace_priors_comparison}) resembles the old one (orange on Fig. \ref{fig:55_iter_Flat_Laplace_priors_comparison}) or not: we found that it is not the case. 
In particular, on Fig. \ref{fig:55_iter_Flat_Laplace_priors_comparison}, we can see that the peak of the {Laplace} posterior is almost at 50 (in terms of value assumed by the density), whereas the maximum value of the half-Laplace prior is 40. {So $25\%$  smaller than the posterior. Moreover, even if} the green posterior is close to a half-Laplace density, {its} tail is shorter. We can thus conclude that the information contained in the data set (and {by} using this methodology) slightly preferers GRT since the posterior is even more peaked toward $m_g=0$. 
\subsubsection{The Bayes Factor}
In {Sec. \ref{sec:nutshell}, Sec. \ref{sec:results_flat_wo_uncertainty} and Appendix \ref{sec:methodMH} }  we explained that, because of the time of computation, it has been chosen to not use the direct {evaluation} of the {$\chi^2$} function but an approximation based on the Gaussian Process. We also used the Gaussian Process as a way to assess the uncertainties of this interpolation and show in Sec. \ref{sec:resultslaplace} that the results were only marginally impacted by the interpolation uncertainties. 
However, we will see that the result of the non-detection presented in Sec. \ref{sec:resultslaplace} can also be sustained by fully integrated and adjusted ephemerides without any interpolation.
In order to assess this point, we computed an estimation of the Bayes Factor using the same set of masses that have been used for the GPR interpolation and for which fully integrated and adjusted INPOP ephemerides have been built, and a $\chi^2$ computed (see Appendix \ref{sec:methodMH}). 
The Bayes factor is a tool used in the context of model selection (see, e.g. \cite{GillJeff2008}). The central notion is that prior and posterior informations should be combined in a ratio that provides evidence of one model specification $M_1$ over an other $M_2$. The Bayes Factor can be interpreted as a quantity saying which model between $M_1$ and $M_2$ represents at the best the observed data set $\mathbf{d}_{\text{obs}}$.
In our case selecting a specific model means to select a specific value of $m_g$.
The Bayesian setup requires a prior distribution for the parameter $m_g$ we are dealing with, that, for us, is $\rho(m_g)$.  
The quantity of interest is then the ratio:
\begin{equation} \label{eq:BF_Definition}
BF= \frac{\pi(M_1 | \mathbf{d}_{\text{obs}} )}{\pi(M_2 | \mathbf{d}_{\text{obs}} )} \times \frac{\rho(M_2)}{\rho(M_1)}.
\end{equation}
In the case of equal priors for $M_1$ and $M_2$ the Bayes Factor equals to the ratio of the likelihoods. For major details about the Bayes Factor and a more extensive explanation the reader is addressed to \cite{GillJeff2008}. The interpretation of the Bayes Factor (BF) is not always easy. We will take as reference the so called Jeffreys' scale {which is a qualitative judgment on the evidence.}Generally speaking, if $BF \gg 1$ it indicates that the weight of Model 1 is greater than the Model 2, and on the other hand if $BF \ll 1$ then Model 2 has more weight. In the zones of the domain for which $BF \simeq 1$ neither Model 1 or Model 2 is predominant. The Bayes Factor is computed with real unapproximated $\chi^2$ and not with interpolated values. As Model 1, we use an extremely small value for the mass of graviton i.e. $m_g \simeq 10^{-40} \times 10^{-23} \; eV c^{-2}$ mimicking GRT; whereas as Model 2, we employ a massive graviton.

\begin{figure*}
    \centering
    \includegraphics[width=.9\linewidth]{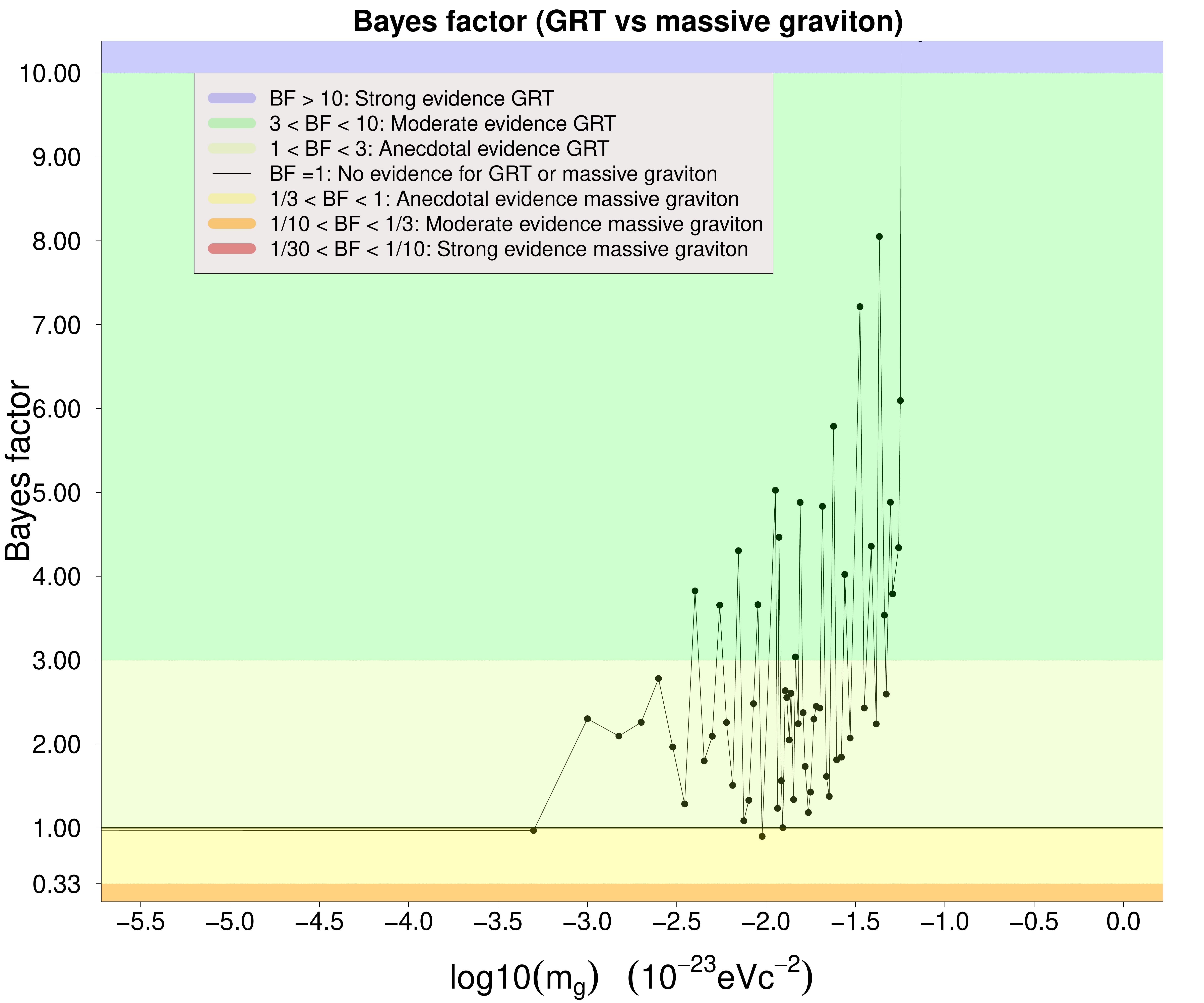}
    \caption{Value of the Bayes Factor as function of $m_g$. Each point corresponds to one INPOP run. On the $x$-axis we put the masses in terms of $\log_{10}(m_g)$. }
    \label{fig:BayesFactor55}
\end{figure*}

On Fig. \ref{fig:BayesFactor55} the Bayes Factor is presented relative to the $\log_{10}$ of the graviton mass. The BF values tend to be above 1 for almost all the values of $m_g$, and it is close to 1 for values of $m_g$ roughly smaller than $10^{-25} eV c^{-2}$. 
The conclusion is also that GRT remains the most likely model ($BF > 10$) up to $m_g = 0.06 \times 10^{-23} \; eV c^{-2} $. For $m_g \leq 0.06 \times 10^{-23} \; eV c^{-2} $, $BF > 1$ almost everywhere, except in some points for which we have $BF\simeq 1$. Below $0.06 \times 10^{-23} \; eV c^{-2} $ the general trend is in any case with anecdotal or moderate evidence towards GRT. On Fig. \ref{fig:BayesFactor55} one can see that we do not obtain any evidence for a massive graviton favored over GRT, for any possible ephemerides with $m_g \neq 0$. 
This conclusion is consistent with what has been discussed in Sec. \ref{sec:resultslaplace} and Sec. \ref{sec:disc_laplace_prior} with less resolution as with the GPR and MH algorithms but in using full estimated $\chi^2$ function without approximations.

\subsection{Comparisons with previous estimates}
\label{sec:discussioncompar}
\subsubsection{Comparison with previous INPOP estimations}
\begin{figure*}
    \centering
    \includegraphics[width=.89\linewidth]{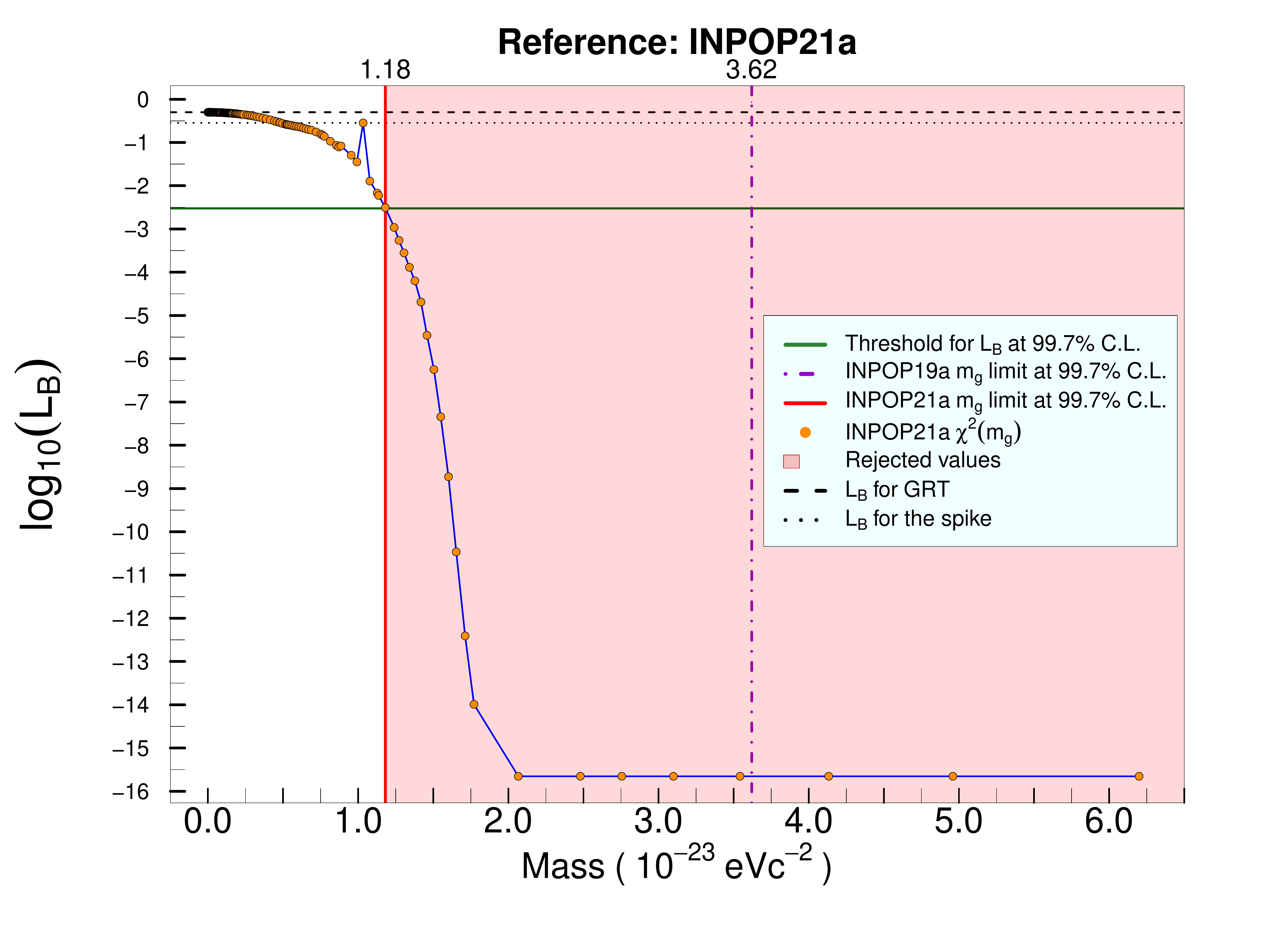}
    \caption{Values of the likelihood $L_B$ as computed in \cite{Bernus2020}. The orange dots represent the $\log_{10}(L_B)$ values computed with INPOP21a. The green horizontal line is a threshold value for $L_B$ at $99.7 \%$ with the same criteria used in \cite{Bernus2020}. The vertical red line is the $m_g$ limit at $99.7 \%$ C.L. with the same criteria adopted in \cite{Bernus2020} but using INPOP21a. The violet vertical line is the  $m_g$ limit at $99.7 \%$ C.L. proposed in \cite{Bernus2020} using INPOP19a. The shaded red zone is the zone of rejected value we would obtain with INPOP21a at $99.7 \%$ C.L. . {The horizontal dashed line represent the $L_B$ value for GRT, whereas the horizontal dotted line represents the $L_B$ value for the spike around $m_g=1.03 \times 10^{-23} \; eVc^{-2}$.  } }
    \label{fig:Likelihood_Bernus_Comparison_v2}
\end{figure*}

\cite{bernus2019, Bernus2020} have given upper bounds for the mass of the graviton obtained with INPOP17b and INPOP19a. In our work are presented a generalization and an improvement of the results given in \cite{bernus2019, Bernus2020}, using INPOP21a. 
By using a more general semi-Bayesian approach, we are showing that GRT is enough for explaining the data, and the massive graviton is not inducing any improvement in the planetary model. %
In order to compare with previous works, we can give an upper limit of the posterior.
In particular in \cite{bernus2019, Bernus2020} the upper bound for the mass with a $99.7 \%$ confidence level is $m_g \leq 3.62 \times 10^{-23} \;  eV c^{-2}$. 
In order to proceed with a comparison, we take the $99.7 \%$ confidence level on GPUR {(see Sec. \ref{sec:results_flat_and_uncertainty} and Table \ref{tab:GPR_300iterations_comparison_55it})} which corresponds to a mass of $m_g \leq 1.01 \times 10^{-24} \;  eV c^{-2} $.
The result is an improvement by 1 order of magnitude in comparison with Bernus et al. \cite{Bernus2020}. It is however interesting to understand if this result is due to the INPOP improvement, or due to the change of methodology.
Hence, we used the formalism proposed in \cite{Bernus2020} with INPOP19a, but using INPOP21a $\chi^2$ values instead. 
\cite{Bernus2020}  computed a "likelihood" interpreted as the probability of a tested theory to be likely. We refer to such a likelihood as $L_B$. The results obtained using INPOP21a are shown on Fig. \ref{fig:Likelihood_Bernus_Comparison_v2}. 
On this Figure, one can see that the INPOP21a $L_B$ value is improved by a factor 3, going from $3.62 \times 10^{-23} \; eVc^{-2}$ to $1.18 \times 10^{-23} \; eVc^{-2}$ for a C.L. at $99.7  \% $. A spike is present on Fig. \ref{fig:Likelihood_Bernus_Comparison_v2} for $m_g=1.03 \times 10^{-23} \;  eV c^{-2} $: this is due to a local minimum in the $\chi^2$ function. But let us stress that, even at this local minimum one has the likelihood $L_B$ (at $m_g = 1.03 \times 10^{-23} \;  eV c^{-2}$) that is still smaller than the likelihood $L_B$ obtained in GRT. Thus, GRT is still the most favourite guess between the two possibilities. The local minimum is also present in the function $m_g \longmapsto \tilde{\chi}^2(m_g)$ used in the MH algorithm: the MCMC process takes into account this local minimum and overcomes it going towards the global minimum at about $m_g \simeq 0$. For further details on the computation of $L_B$ see \cite{Bernus2020}. We can conclude that both the new model, and the new observations introduced with INPOP21a induce a factor 3 improvement relative to \cite{Bernus2020} determination on $m_g$. The MH algorithm implemented here goes even further and explores a zone close to $m_g=0$.

\subsubsection{Comparison with LIGO-Virgo-KAGRA estimations}
The {LIGO-Virgo-KAGRA} collaboration presents an updated bound of the mass of the graviton $m_g$ at $90 \%$ confidence level, that is $m_g \leq 1.27 \times 10^{-23} eV c^{-2}$ in \cite{LIGOVirgo_testsGRT_12_2021}. The posterior then obtained from the MCMC in our work seems to {give a 1 order of magnitude better constraint. However, it is important to stress that}
the two studies are perfectly complementary because they focus on different aspects of the massive gravity phenomenology (radiative versus orbital), and also use totally different observations (gravitational waves versus astrometry in the Solar System). {In particular, decoupling mechanisms (in some specific theories of massive gravity) could in principle suppress one aspect of the massive gravity phenomenology and not the other, such that, from a phenomenological point of view, it will always be important to probe both aspect of the massive gravity phenomenology.}

\subsection{Improvement with the BepiColombo mission}

\begin{figure*}
    \centering
    \includegraphics[width=.89\linewidth]{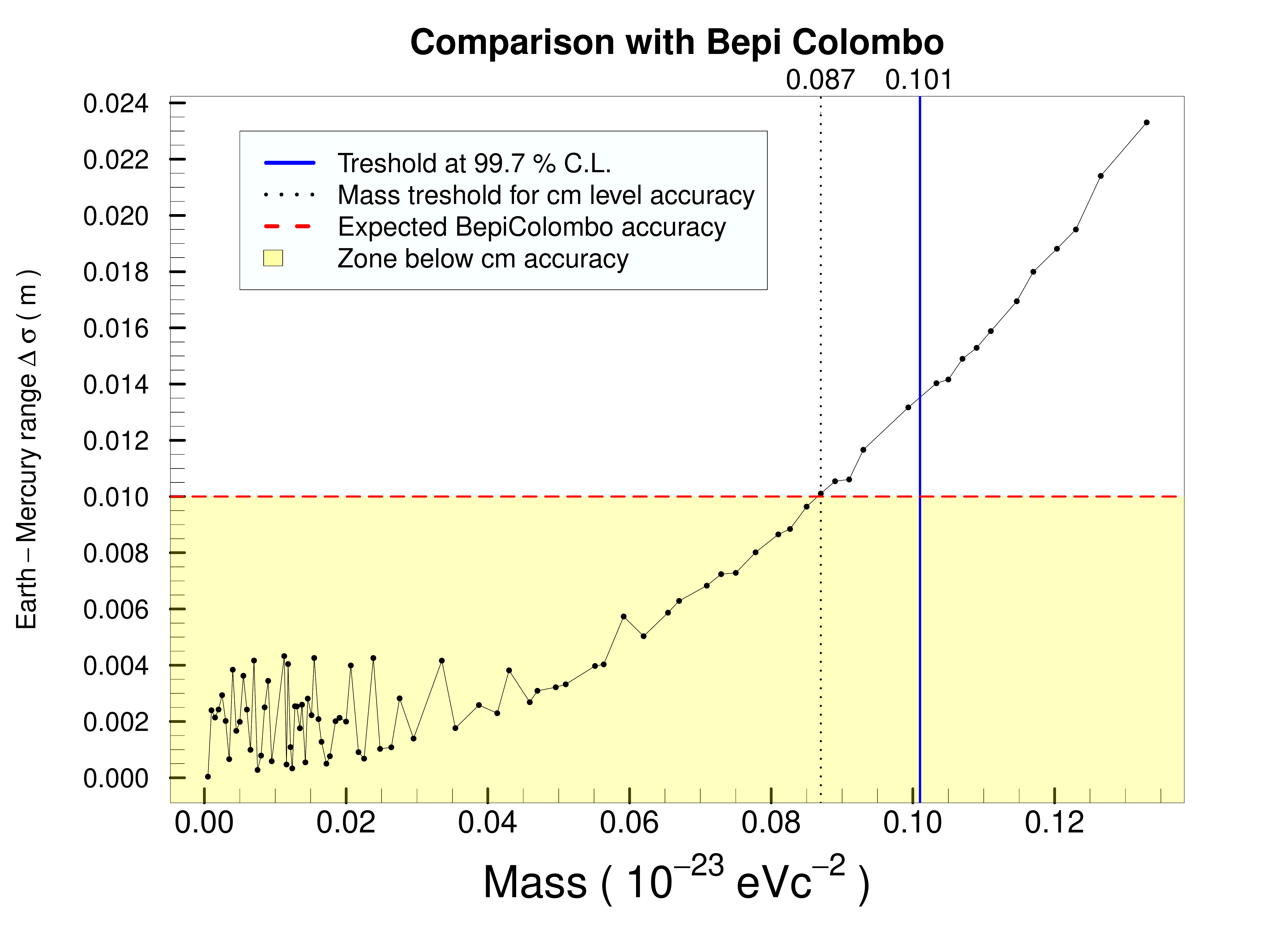}
    \caption{Standard deviations of the differences on Earth-Mercury distances ($\Delta \sigma$) between INPOP21a and massive graviton ephemerides as function of the massive graviton $m_g$. The red dashed horizontal line represents the BepiColombo measurement uncertainty whereas the yellow area represents the zone below the 1 cm accuracy. The black dotted vertical line gives the smallest detectable $m_g$ value.}
    \label{fig:Comparison_with_BC}
\end{figure*}

The BepiColombo mission will arrive at Mercury in 2026 for at least 2 years of close circular orbit about the planet.
Thanks to the MORE radio science experiment, measurements of the Mercury to Earth distances will be obtained with an accuracy of about 1~cm in KaKa-Band \cite{2021SSRv..217...21I}. 
Based on such an accuracy, we suppose for our simulations, a daily acquisition of range tracking data  during a period of 2.5 years, from 2026 to 2028.5 \citep{thor20} and in using INPOP21a as reference solution in GRT framework. In order to assess if the MORE range data will be sensitive to the mass of the graviton and up to which level these measurements can help to improve the results of this work, we computed the differences for the Earth-Mercury distances during the 2 years of the MORE observation period between INPOP21a and ephemerides built and fitted with different values of the graviton mass $m_g$. As it has been already discussed, a given $m_g$ value fixed in the computation of the ephemerides yields to a perturbation of the orbits, and then on the residuals, with respect to ephemerides in which GRT is assumed. Quantifying such a perturbation provides a way to see if data from future interplanetary missions may play a role in the $m_g$ limit determination, keeping in mind that after fit of the perturbed ephemeris, most of the perturbation will be absorbed by other parameters (see discussion about correlations in i.e. \cite{bernus2019} or \cite{LRR22}). As the expected BepiColombo accuracy for the range is at 1 cm-level, one can expect that for perturbations below this threshold of 10$^{-2}$~m would not be detectable by the MORE measurements. On Fig. \ref{fig:Comparison_with_BC}, the standard deviations $\Delta \sigma$ of the Earth-Mercury distance perturbations  are presented as function of the massive graviton $m_g$ considered. We can see that the smallest $m_g$ that might produce a significant perturbation is roughly $m_g=0.087 \times 10^{-23} eVc^{-2}$, which is only $13.8 \%$ below the bound provided in Sec. \ref{sec:results_flat_and_uncertainty}. 
Because of correlations between parameters, we are not considering this limit as the one that would minimize the future INPOP $\chi^2$ including the MORE data, but more as a minimum threshold below which the mass of the graviton will not be detectable by the BepiColombo radio science experiment.
As a consequence, we expect a marginal improvement on the graviton mass from the BepiColombo mission as the threshold of detection with MORE will be only $13.8 \%$ smaller than the limit given  in Sec. \ref{sec:results_flat_and_uncertainty}

\section{Conclusion}

We have presented our work on the use of MCMC algorithm and INPOP in order to get an improvement of the detection limit of the mass of the graviton $m_g$ using the Solar System dynamics as arena. 
A key strength of the present study is to include the $m_g$ contribution in terms of accelerations and light times computation within the full dynamics of the Solar System. Moreover, by considering a semi-Bayesian approach, and using MCMC and MH algorithm, we avoid correlations between $m_g$ and other INPOP astronomical parameters.
We used the GPR to obtain an approximation of the $\chi^2$ ready to use within the MH algorithm and to asses the uncertainty of approximation afterwards. 
From the posterior obtained, we can give an upper bound of $m_g \leq 1.01 \times 10^{-24} \; eVc^{-2}$ at $99.7 \%$ C.L. {(resp. $\lambda_g \geq 122.48 \times 10^{13} \; km$) }, including approximation uncertainty, and we had shown with a change of the prior (from flat to  half-Laplace) that no significant information is detectable in planetary ephemerides for masses smaller than this limit. 
Observations from the BepiColombo mission will provide new data that, according to our analysis, should not lead to a major improvement on the mass of the graviton.

 \section{Acknowledgments}
VM was funded by CNES (French Space Agency) and UCA EUR Spectrum doctoral fellowship. This work was supported by the French government, through the UCAJEDI Investments in the Future project managed by the National Research Agency (ANR) under reference number ANR-15-IDEX-01. The authors are grateful to the OPAL infrastructure and the Université Côte d’Azur Center for High-Performance Computing for providing resources and support. VM and AF thank  A. Chalumeau, C. Twardzik, S. Babak and L. Bigot for their useful inputs and discussions.

\bibliography{Bibliography_natbib,global}

\appendix

 \section{Method}
 \label{sec:method}
 
 In this section are given the most important aspects of the method for the Metropolis-Hasting (MH) algorithm associated with an interpolation {of the $\chi^2$ function} and the uncertainty assessments based on Gaussian Process (GP). Complementary information and fully detailed implementation, as well as validation tests, can be found in \cite{2023arXiv230305298M}.

 \subsection{{Metropolis-Hastings algorithm and Gaussian Process Regression}}
 \label{sec:methodMH}

 In the past years there have been already some attempts to deal with the problem of high correlations among parameters inside the INPOP planetary ephemerides fit,
as in the case of the determination of asteroid masses (see \cite{10.1093/mnras/stz3407}). For testing alternative theories and thus assessing threshold values for the violation of GRT, \cite{Fienga-2015} had tested genetic algorithm approaches for identifying intervals of values for parameters such as PPN, $\beta$, $\gamma$, the Sun oblateness $J_2$ and secular variations of the gravitationnal mass of the Sun $\frac{\Dot{\mu}}{\mu}$, with which planetary ephemerides can be computed and fitted to the observations with a comparable accuracy than the ephemerides built in GRT.

Keeping in mind the problem of correlation between planetary ephemerides and GRT parameters, we propose a new procedure with a semi-Bayesian approach to test a possible deviation from GRT in a particular case: we investigate the posterior probability distribution of a possible non-zero mass of the graviton $m_g$ employing MCMC techniques.

Our approach is semi-Bayesian in the sense that only the mass of the graviton is actually sampled with the Metropolis-Hastings (MH) algorithm procedure, {the INPOP astronomical parameters being fitted with a least square procedure}. We follow here the algorithm already used by \cite{bernus2019,Bernus2020,2022IAUS..364...31F}: for a fixed value of $m_g$, we integrate the motion of the planets with INPOP and we fit to planetary observations, the astronomical parameters listed in Sec. \ref{sec:methodinpop} in using the least square iterative procedure described in \cite{10.1093/mnras/stz3407}. We then obtain a fully fitted ephemeris built for a fixed value of $m_g$. 
%
%
{The fitted ephemerides are necessary in order to compute the likelihood, essential part of any MCMC procedure.} Generally speaking, the Metropolis-Hastings algorithm is one of the first Markov Chain Monte Carlo methods developed, providing a sequence of random samples drawing them from a given probability distribution. {The probability distribution from which we draw our samples is the posterior, taking into account the Solar System dynamics and all the observations used in INPOP. The MH algorithm is based on the idea that the drawings are done sequentially and according to a random acceptance process. The outcome of the algorithm is a sequence of random samples from the posterior. Indeed, this sequence is a Markov Chain, and its equilibrium distribution is the posterior itself. At each step of the algorithm, one new element of the sequence is computed and proposed, being accepted or rejected. The acceptance/rejection is random, according to a certain probability (changing at each step) based on the last accepted element of the sequence and the likelihood of the last element and of the new candidate element.} We will not provide a proof of convergence of the method since it is out of the goal of our work and it can be found easily in \cite{RobertCasella2004} and  \cite{brooks2011handbook}. For a detailed overview on the MCMC method see \cite{brooks2011handbook} and \cite{RobertCasella2004}. The details about the specific MH implementation adopted for this work can be found in \cite{2023arXiv230305298M}.  
The MH algorithm will play {a role} only on the selection of the parameter of interest (i.e. the mass of the graviton $m_g$) whereas the 402 other INPOP parameters are obtained by regular least squares. This hybrid approach has been chosen because, from one side, it would be very costly in term of time of computation to explore the full parameter space with a MH algorithm, {and, on the other side, this is a way to avoid correlation of $m_g$ with the other INPOP parameters.}
Similarly to \cite{Bernus2020}, {the $\chi^2(m_g)$ computed will be estimated setting fixed the value of $m_g$ and fitting the remaining $\mathbf{k}$ parameters (see Eq. \eqref{normalized_chi2_form1}) with the full INPOP adjustment}. {Differently} from \cite{Bernus2020}, {we provide a posterior probability distribution for} $m_g$, {and it is done with} the MH algorithm.
\\
Finally, we approximate the {$\chi^2$} function {with} $m_g \longmapsto \tilde{\chi}^2 (m_g)$ considering a set of values $ S = \{ \left( \overrightarrow{m}_g, \chi^2(\overrightarrow{m}_g) \right) \} $ and interpolating among the points of the set $S$. $\overrightarrow{m}_g$ gathers values of the masses for which we are going to compute the actual corresponding $\chi^2$ unapproximated.
The interpolation $m_g \longmapsto \tilde{\chi}^2 (m_g)$ has been built in exploiting a Gaussian Process among the points of the set $S$.
The Gaussian Process Regression (GPR) is a method to predict a continuous variable as a function of one or more dependent variables, where the prediction takes the form of a probability distribution (see e.g. \cite{Strub:2022upl, rasmussen2005gaussian}). A full description of the method can be found in \cite{2023arXiv230305298M} and the reader is referred to \cite{rasmussen2005gaussian} for a more detailed discussion on the Gaussian Processes and to \cite{GPfit2015} for the documentation of the package \textbf{GPfit} that we used to produce our GPR. The notation used for the general GPR description is similar to what is found in \cite{Strub:2022upl, rasmussen2005gaussian}. {To compute a Gaussian Process Regression among a given set of points, an function $m$ (in GPR jargon the \textit{mean function}) and a function $K$ (in GPR jargon the \textit{kernel function}) are necessary.} In our case the $m$ function is the Best Linear Unbiased Predictor (BLUP), as described in \cite{GPfit2015} and \cite{Robinson1991}. As kernel we used a standard exponential kernel such as \[ K(x, x') = \exp \left( - \frac{|x - x'|^{\alpha}}{ \ell^2} \right) \]  for suitable $\ell$ and $\alpha$ parameters to tune.
The points we are interpolating with the GPR (in GPR jargon the \emph{observations}, see \cite{rasmussen2005gaussian} ) are considered as noise-free, since they are computer simulations. We refer the reader to \cite{Sacks1989} for further details about this assumption. 
However, as we are using the GPR for interpolating $\chi^2$ values that will be seen as forward model outcome by the MCMC, it is interesting to consider the uncertainty of this interpolation (see \cite{Sacks1989} and \cite{Santner2003}) for the interpretation of the MCMC results, {propagating this uncertainty in the posterior sampled by the MH algorithm}.

\begin{figure}[!ht]
\centering
  \includegraphics[width=\linewidth]{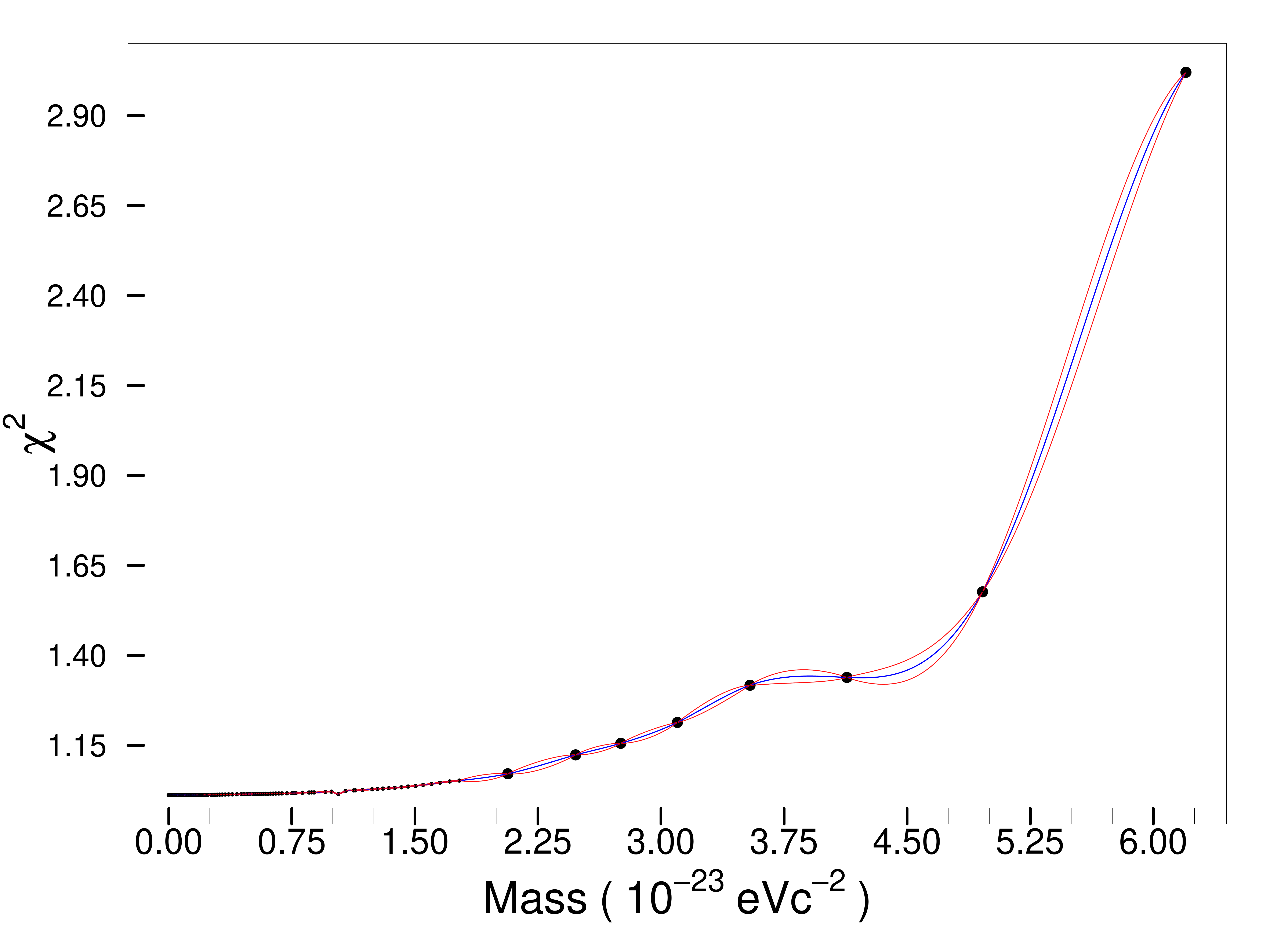}
   \caption{Plot of the function $ m_g \longmapsto \tilde{\chi}^2(m_g)$. The dots correspond to $\chi^2$ value estimated with INPOP and not extrapolated. the red lines correspond to the $2-\sigma$ uncertainties provided by the GPR.}
    \label{fig:GPR_mass_vs_chi2}
\end{figure}

\begin{figure*}[!ht]
\centering
  \includegraphics[width=\linewidth]{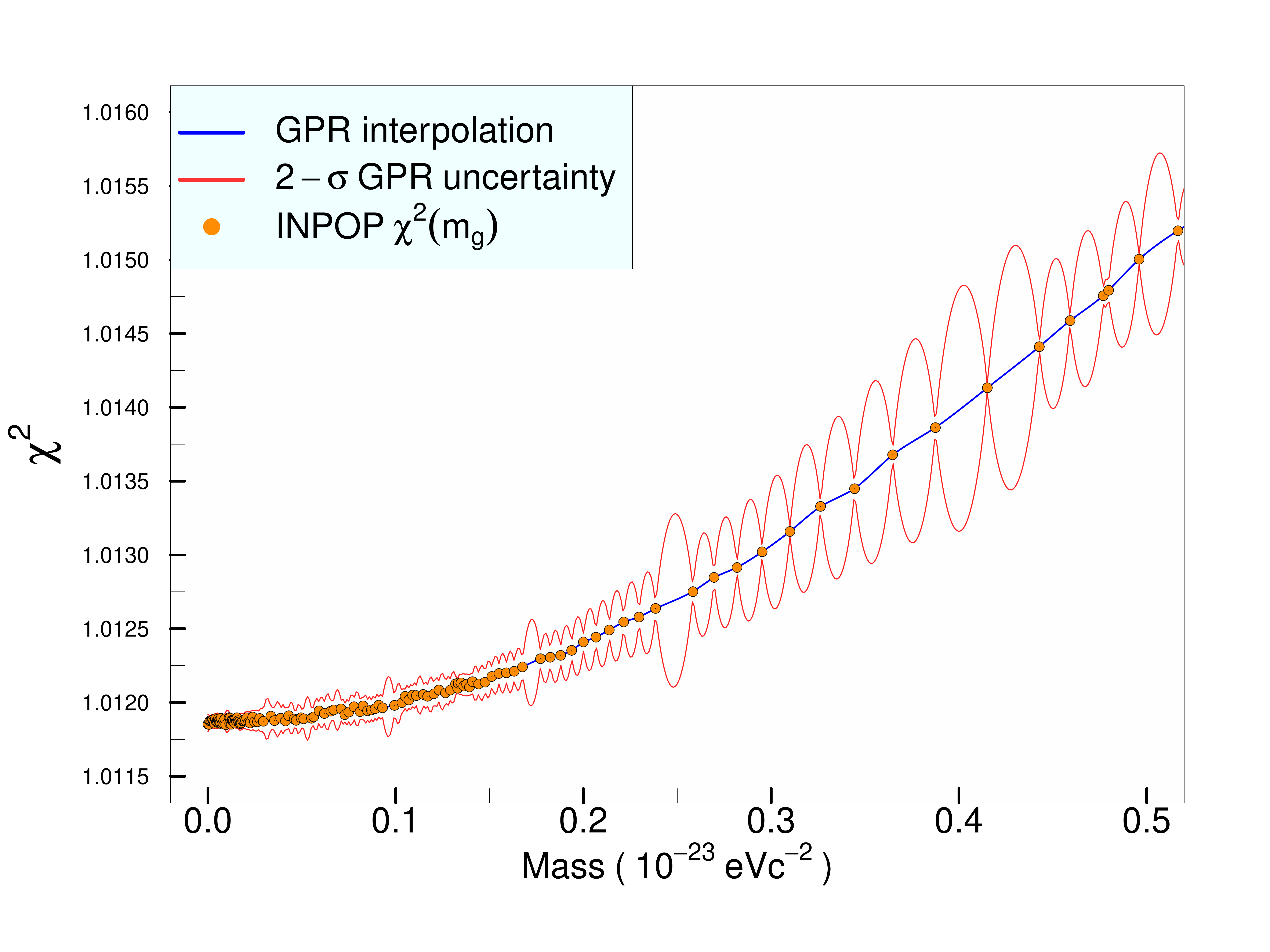}
    \caption{ Plot of the function $ m_g \longmapsto \tilde{\chi}^2(m_g)$ made up with INPOP $\chi^2$. The GPR interpolation is plotted in blue among the points $\chi^2(m_g)$ computed with the full INPOP (orange dots). The red lines represent the $2-\sigma$ uncertainty (on the interpolation) provided by the GPR. }
    \label{fig:GPR_mass_vs_chi2_zoom3}
\end{figure*}

On Fig. \ref{fig:GPR_mass_vs_chi2} we see the evolution of the $\chi^2$ as a function of $m_g$ as well as the GPR outcome. The black dots indicated the $\chi^2$ obtained with fully integrated and converged ephemerides when the blue line represents the mean value obtained from GPR (see details in \cite{2023arXiv230305298M}). Moreover the red lines indicate the estimates of uncertainty provided by the GPR at $2 \sigma$ level. In particular the graphs of the two red lines are $m_g \longmapsto \tilde{\chi}^2(m_g) \pm 2\tilde{\sigma}(m_g) $. We see that $m_g \longmapsto \tilde{\sigma}(m_g)$ is zero or close to zero when $m_g$ is used to compute $\chi^2(m_g)$ from INPOP. This is due to the assumption that the $\chi^2( \overrightarrow{m}_g )$ are computed with zero noise as these values are directly obtained from INPOP construction. {On Fig. \ref{fig:GPR_mass_vs_chi2_zoom3} a zoom of Fig. \ref{fig:GPR_mass_vs_chi2} is provided, on the final interval of interest of the MH algorithm}. 

\subsection{Gaussian process and uncertainty assessments}
 \label{sec:methodsig}

 One of the advantages about GPR, is that, in principle, we can consider the value of $\tilde{\chi}^2(m_g)$ as a normal random variable with given average and standard deviation provided as outcome of the GPR. 
As we said previously the estimator in our case is BLUP and we call it $m_g \longmapsto \tilde{\chi}^2(m_g)$. 

 We would like also to exploit these confidence intervals and the fact that, due to the hypothesis of Gaussian Process and the assumptions we did, Eq. \eqref{grayline_form2} holds:
\begin{equation}\label{grayline_form2}
\forall m_g, \quad \chi^2(m_g) \sim \mathcal{N}(\tilde{\chi}^2(m_g), \tilde{\sigma}^2(m_g)).
\end{equation} 

Based on Eq. \eqref{grayline_form2}, we can produce a \textit{perturbation} of the interpolation $m_g \mapsto \tilde{\chi}^2(m_g)$, that, for sake of clarity, we are going to call Gaussian Process Uncertainty Estimation (GPUE). The underlying idea of GPUE being that, for each $m_g$ of the domain, you can consider the corresponding $\chi^2(m_g)$ as outcome of a random draw of a normal random variable according to the uncertainty of the GPR, i.e. following Eq. \eqref{grayline_form2}. 

\begin{figure*}
\centering
  \includegraphics[width=\textwidth]{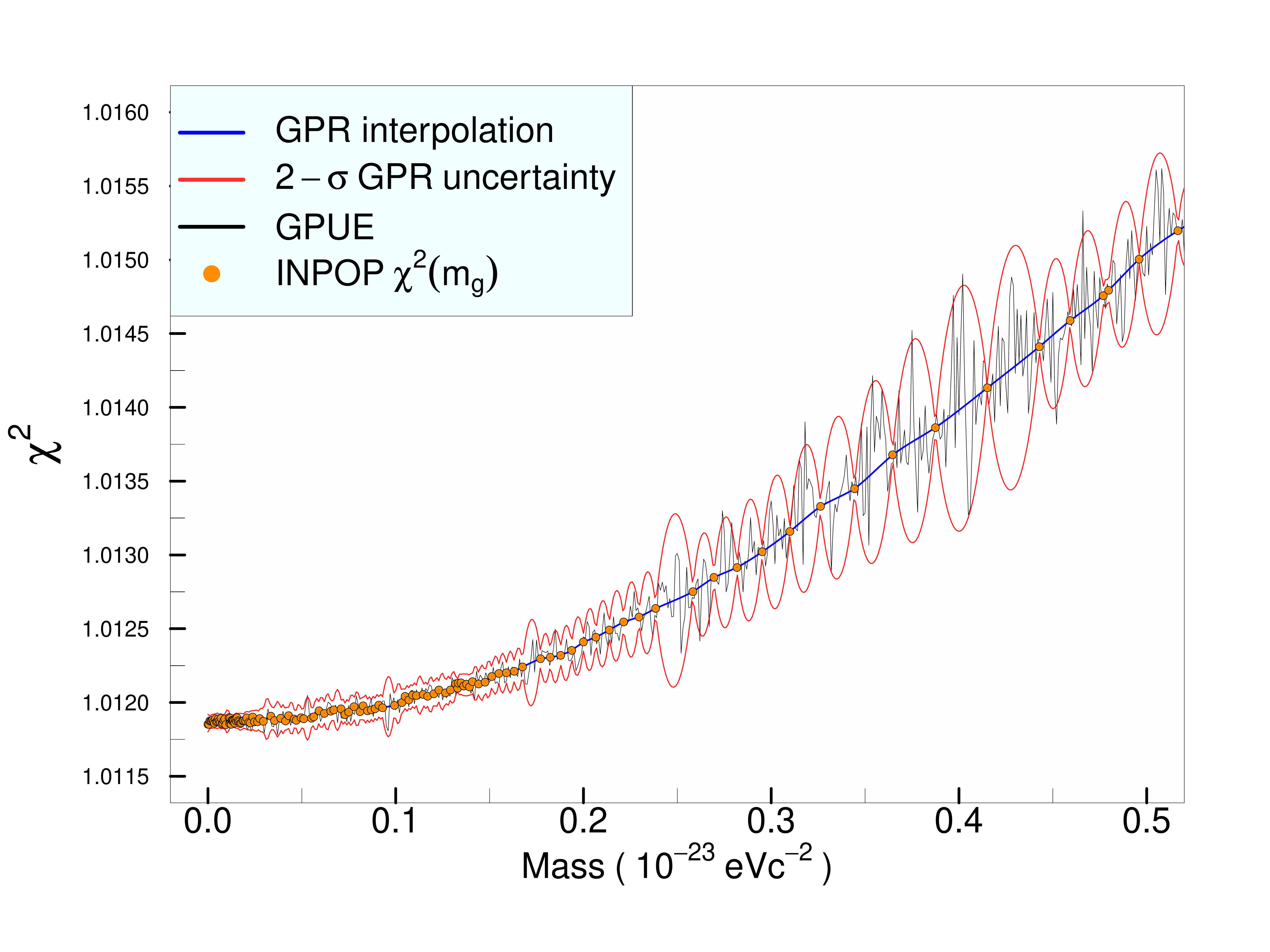}
    \caption{ Plot of the function $ m_g \longmapsto \tilde{\chi}^2(m_g)$ made up with INPOP $\chi^2$ (orange dots). The GPR interpolation is plotted in blue among the points $\chi^2(m_g)$ computed with the full INPOP (orange dots). The red lines represent the $2-\sigma$ uncertainty (on the interpolation) provided by the GPR. The black like is one possible GPUE (in other words, the GPR interpolation perturbed at each point of the domain). The plot is zoomed on a portion of the domain.}
    \label{fig:GPR_only_plot_IT55_zoom3_nogrid_pert}
\end{figure*}

In Eq. \eqref{grayline_form2} $\tilde{\sigma}(m_g)$ is the value provided by the GPR to use as standard deviation when we consider $\chi^2(m_g)$ as a normal random variable.
{
 Let's indicate as 
 \begin{equation}\label{grayline_form1} 
 m_g \longmapsto \breve{\chi}^2(m_g), 
 \end{equation}
one \emph{estimation} of the GPR with its own uncertainty.
On Fig. \ref{fig:GPR_only_plot_IT55_zoom3_nogrid_pert} the black line represent one GPUE. For a detailed description of how to build such a GPUE, the reader is addressed to \cite{2023arXiv230305298M}.
}
 
By definition $m_g \longmapsto \breve{\chi}^2(m_g)$ is locally defined as a realization of a normal random variable. 
{In order to propagate the GPR uncertainty in the MCMC, we are going to use the GPUE. We compute several GPUEs, all different, whereupon for each of the maps $m_g \longmapsto \breve{\chi}^2(m_g)$ produced, we run separately a Markov Chain using the MH algorithm.}  Doing so we are running MH algorithm to produce a Markov Chain on a noisy version of the nominal interpolation obtained with GPR. {All the Markov Chains obtained are joint together, producing the (blue) posterior presented on Fig. \ref{fig:55_iter_300Pert_GPR_histogram_comparison}.} This method is an attempt to assess the \emph{uncertainty of interpolation} that we have. We call Gaussian Process Uncertainty Realization (GPUR) such a final posterior.




\end{document}